# Destructive Controlled-Phase Gate Using Linear Optics

S.U. Shringarpure and J.D. Franson
*Physics Department, University of Maryland Baltimore County, Baltimore, MD 21250 USA*

**Abstract**: Knill, Laflamme, and Milburn [Nature **409**, 46 (2001)] showed that linear optics techniques could be used to implement a nonlinear sign gate. They also showed that two of their nonlinear sign gates could be combined to implement a controlled-phase gate, which has a number of practical applications. Here we describe an alternative implementation of a controlled-phase gate that only requires the use of a single nonlinear sign gate. This gives a much higher average probability of success when the required ancilla photons are generated using heralding techniques. This implementation of a controlled-phase gate destroys the control qubit, which is acceptable in a number of applications where the control qubit would have been destroyed in any event, such as in a postselection process.

## I. Introduction

A controlled-phase gate produces a phase shift $\phi$ when the control and target qubits both have a logical value of 1. This is a very useful operation since it is a universal gate for quantum computation when combined with single-qubit operations [1]. It can also be used to create Schrodinger cat states [2], to perform nonlocal quantum interferometry with violations of Bell's inequality [3,4], and to implement complete Bell state measurements in quantum teleportation [5,6], for example.

Knill, Laflamme, and Milburn (KLM) [7] showed that linear optics techniques could be used to implement a nonlinear sign gate. They also showed that two of their nonlinear sign gates could be combined to implement a controlled-phase gate. In this paper, we propose an alternative implementation of a controlled-phase gate that only requires a single nonlinear sign gate. Since each operation of a nonlinear sign gate requires an ancilla photon, our approach requires one less ancilla photon than earlier approaches [7,8]. This gives a higher average probability of success when the required ancilla photons are generated using down-conversion and heralding techniques. The increased probability of success comes at the expense of destroying (erasing) the control qubit.

Logic gates in which the control qubit is destroyed have been used in a number of previous applications. For example, a destructive Controlled-NOT (CNOT) gate can be combined with a quantum encoder to implement a non-destructive CNOT gate [9,10]. The same devices can be used to implement fusion gates that allow the construction of a cluster state [11]. As another example, Bell's inequality can be violated in nonlocal interferometer experiments in which a controlled-phase shift is combined with homodyne measurements [4]. The control qubit is destroyed in a postselection process in experiments of that kind, which allows the use of the controlled-phase gate described here.

The remainder of the paper is organized as follows. Section II describes two realizations of a nonlinear sign gate. The controlled-phase gate proposed by KLM is briefly reviewed in Section III. Section IV proposes an alternative implementation of a controlled-phase gate that only requires a single nonlinear sign gate. The performance of the destructive controlled-sign gate is compared with that of the KLM gate in Section V. A technique that allows a controlled-phase gate to operate on inputs containing a large number of photons, such as a coherent state, will be described in Section VI. A summary and conclusions are provided Section VII.

## II. Nonlinear sign gates

The nonlinear sign gate shown in Fig. 1 is the basic building block of the KLM approach to linear optics quantum computing [1]. The input state $|\psi_{in}\rangle$ is limited to at most two photons. The operation of the nonlinear sign gate is then defined by

$$|\psi_{in}\rangle = \alpha|0\rangle + \beta|1\rangle + \gamma|2\rangle \rightarrow \alpha|0\rangle + \beta|1\rangle - \gamma|2\rangle, \quad (1)$$

where $\alpha$, $\beta$, and $\gamma$ are complex constants. The only effect of the nonlinear sign gate is to reverse the sign of the two-photon amplitude, which is similar to the effects of a nonlinear Kerr medium [12].

The KLM nonlinear sign gate utilizes three beam splitters, one ancilla photon, and postselection based on the output of two single-photon detectors, as shown in Fig. 1.

The gate applies a nonlinear phase shift of $\pi$ as in Eq. (1) for an appropriate choice of beam splitters and linear phase shifters as shown in Fig. 1. Other choices of the parameters can also be used to implement a nonlinear phase shift of $\pi/2$, for example [7]. There have been several proposals to enhance the success rate of this gate at the expense of adding more resources [13,14] or vice-versa [15].

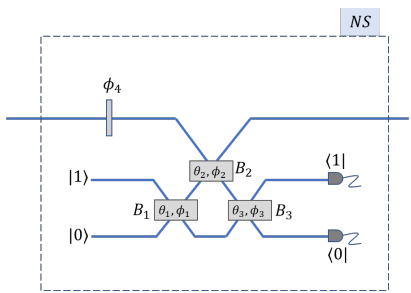

FIG. 1. The KLM nonlinear sign gate. An input state of the form $|\psi\rangle = \alpha|0\rangle + \beta|1\rangle + \gamma|2\rangle$ gives an output state $\alpha|0\rangle + \beta|1\rangle - \gamma|2\rangle$ for an appropriate choice of the transmission coefficients of the three beam splitters $B_1$, $B_2$, and $B_3$, along with a fixed phase shift $\phi_4$. The results are heralded on the presence of a single photon in one of the two single-photon detectors.

Costanzo et al. [12] proposed an alternative implementation of a nonlinear sign gate that is shown in Fig. 2. As illustrated in the upper part of the figure, the device produces a coherent superposition of photon subtractions that occur either before or after a photon addition. The operation of the gate can be intuitively understood from the commutation relation $[\hat{a},\hat{a}^\dagger]=1$. This gate can be implemented using a down-conversion crystal with heralding to produce the photon addition, with photon subtraction occurring either at the first beam splitter $B_1$ or the second beam splitter $B_2$. Heralding on the output of beam splitter $B_3$ ensures that there is a fixed phase relationship between the two ways in which the photon subtraction can occur. The final state in this approach undergoes a noiseless amplification [12] in addition to the nonlinear sign shift. If necessary, this can be compensated using noiseless attenuation [16,17].

Our destructive controlled-phase gate could be implemented using either the KLM nonlinear sign gate or the alternative implementation shown in Fig. 2. Our goal is to implement a controlled phase shift using only linear optical elements, whereas the approach shown in Fig. 2 is based on the use of a nonlinear crystal. As a result, we will assume that the KLM approach is used for the nonlinear sign gate throughout the rest of this paper.

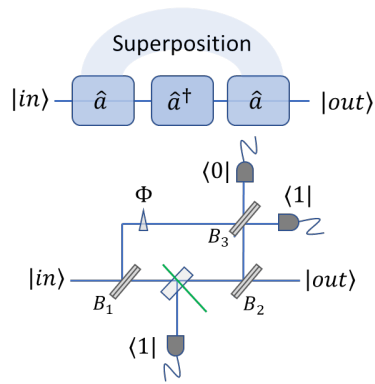

FIG. 2. Alternative nonlinear sign gate suggested by Costanzo et al. [12]. A superposition of $\hat{a}\hat{a}^\dagger$ and $\hat{a}^\dagger\hat{a}$ operations is implemented using photon subtraction that occurs either at the first beam splitter $B_1$ or at the second beam splitter $B_2$. These operations cannot be distinguished when a single photon is detected in one of the outputs of the third beam splitter $B_3$. Photon addition is implemented in between $B_1$ and $B_2$ with the aid of a heralding signal from a down conversion process. A variety of nonlinear phase shifts can be achieved by adjusting the reflectivities of the three beam splitters along with an additional phase shift $\Phi$.

## III. KLM controlled-phase gate

The controlled-phase gate suggested by KLM is shown in Fig. 3. Dual-rail encoding is used for both qubits, and the two paths corresponding to a logical value of 1 are fed into a 50/50 beam splitter. Both outputs of the first beam splitter are passed through a nonlinear sign gate, after which they are recombined on a second beam splitter to form the output of the device.

The operation of this device can be understood as being due to Hong-Ou-Mandel interference [18] at the first beam splitter. If both qubits have a logical value of 0, then no photons pass through the nonlinear sign gates and the device has no effect. If only one qubit has a logical value of 1, then a single photon passes through one of the nonlinear sign gates, which also has no effect. But if both qubits have a value of 1, then both of them will emerge in the same path after the first beam splitter as in the Hong-Ou-Mandel interferometer. In that case, one of the nonlinear sign gates will apply a phase shift of $\pi$ as desired. The second beam splitter can be viewed as implementing the inverse of the Hong-Ou-Mandel interferometer with a single photon emerging in each path.

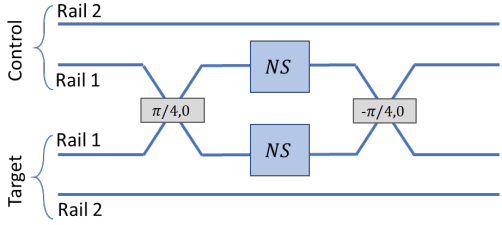

FIG. 3. Controlled phase gate proposed by KLM [7]. Dual-rail encoding is combined with Hong-Ou-Mandel interference at the first beam splitter to apply a phase shift of $\pi$ if both qubits have a logical value of 1. Two nonlinear sign gates labelled NS are required.

Other nonlinear phase shifts, such as $\phi = \pi/2$, can be produced by adding fixed phase shifts and varying the reflectivities of the beam splitters in the nonlinear sign gate from Fig. 1. E. Knill [8] has also described a somewhat a different implementation of a controlled-phase gate that also requires two ancilla photons as a resource.

## IV. Destructive controlled-phase gate

An alternative implementation of a controlled-phase gate that only requires a single nonlinear sign gate is shown in Fig. 4. In this case, we assume that a dual-rail encoding is used for the control qubit while a single-rail encoding is used for the target qubit. The two paths for the control qubit are incident on beam splitters $B_1$ and $B_2$, whose outputs are postselected on the absence of a photon to produce a photon addition at one of the two beam splitters. The path representing a logical value of 1 for the control qubit is assumed to be on the left-hand side of the figure, where it passes through beam splitter $B_1$. A nonlinear sign gate is placed between the two beam splitters, after which beam splitter $B_3$ is used to subtract a photon.

The initial states $|\psi_T\rangle$ and $|\psi_C\rangle$ for the target and control qubits, respectively, will be denoted by

$$|\psi_T\rangle = \alpha|0_T\rangle + \beta|1_T\rangle$$
$$|\psi_C\rangle = \gamma|0_C\rangle + \delta|1_C\rangle, \qquad (2)$$

where $\alpha$, $\beta$, $\gamma$, and $\delta$ are complex constants. Here $|0_T\rangle$ and $|1_T\rangle$ represent the state of the target qubit containing zero or 1 photons, while $|0_C\rangle$ and $|1_C\rangle$ correspond to the dual-rail encoded states of the control qubit.

The basic idea behind the operation of the gate is illustrated in the upper part of Fig. 4. If the control qubit has a logical value of 1, the photon addition occurs first and the state $|\psi'_T\rangle$ that passes through the nonlinear sign gate will contain two photons if the target qubit also has a logical value of 1. In that case, the nonlinear sign gate would produce a phase shift of $\pi$, after which the photon subtraction at beam splitter $B_3$ would restore the target qubit to its original number of photons. In all other cases, the state $|\psi'_T\rangle$ passing through the nonlinear sign gate would contain at most a single photon and no phase shift would be applied.

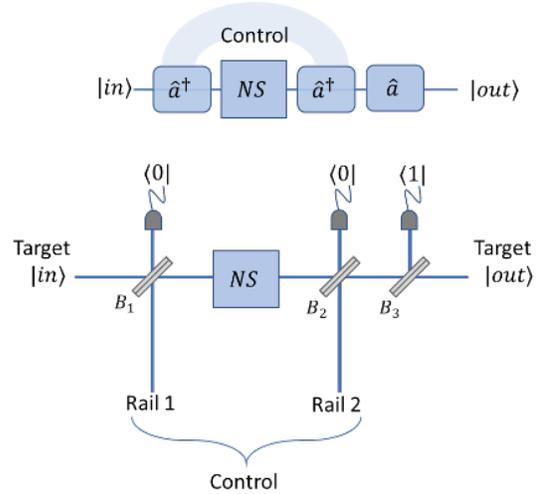

FIG. 4. Implementation of a destructive controlled-phase gate that only requires a single nonlinear sign gate labelled NS. If the control qubit has a logical value of 1, it produces a photon addition at beam splitter $B_1$. If the target qubit also has a logical value of 1, two photons will then pass through the nonlinear sign gate and produce a phase shift of $\pi$. In all other cases, at most a single photon passes through the nonlinear sign gate and there is no effect on the state of the system. A photon subtraction at beam splitter $B_3$ restores the original number of photons to the target qubit. The events are heralded on the outputs shown in three single-photon detectors. The detector in one of the output ports of beam splitter $B_3$ is assumed to be a photon-number resolving detector.

The transmission coefficients for the three beam splitters will be denoted by $t_1$, $t_2$, and $t_3$, while the corresponding reflection coefficients will be denoted by $r_1$, $r_2$, and $r_3$. If we apply the usual beam splitter transformation with a factor of $i$ on reflection, the unnormalized state of the system at the output can be shown to be given by

$$|\psi'\rangle = \gamma r_2 r_3 \left[\alpha|0_T\rangle + \beta(2t_1 t_2 t_3)|1_T\rangle\right] \\ + \delta r_1 r_3 t_2 \left[\alpha|0_T\rangle - \beta(2t_1 t_2 t_3)|1_T\rangle\right]. \quad (3)$$

This state can be put into the desired form by choosing the values of the transmission coefficients such that $2t_1 t_2 t_3 = 1$ and $r_2 = r_1 t_2$. Eq. (3) then reduces to

$$|\psi'\rangle = r_2 r_3 \left[\gamma\left(\alpha|0_T\rangle + \beta|1_T\rangle\right) + \delta\left(\alpha|0_T\rangle - \beta|1_T\rangle\right)\right]. \quad (4)$$

The probability of success is given by $\langle\psi'|\psi'\rangle$, which will depend on the value of the probability amplitudes in the initial state, as discussed in the next section.

Eq. (3) gives a controlled phase shift of $\phi = \pi$ using the parameters described above. Other nonlinear phase shifts can be produced using different parameters in the nonlinear sign gate.

### V. Performance comparison

The probability of success for the destructive controlled-phase gate proposed here will be compared to that of the original KLM controlled-phase gate in this section. The fidelity of both gates depends on the efficiency of the single-photon detectors used in the heralding process, and those efficiencies will also be compared.

One measure of the probability of success is to assume that the necessary ancilla photons are available with 100% probability and then calculate the intrinsic probability of success associated with the gate itself. But in many applications, the relevant probability of success would combine the intrinsic probability of success with the probability of generating the required ancilla photons using down-conversion and heralding techniques. Single photons can be generated using down-conversion with a very high fidelity, for example, which is essential in meeting the threshold for error correction.

We will first consider the probability of success for a controlled-phase gate with $\phi = \pi$. As was noted in the previous section, Eqs. (3) and (4) will give the desired result if we choose $2t_1 t_2 t_3 = 1$ and $r_2 = r_1 t_2$, but those two equations do not completely determine the value of all three transmission coefficients. Figure 5 shows the solutions for $t_1$ and $t_2$ as a function of $t_3$; the solutions only exist for $t_3 > 0.5$. It can be shown that the maximum probability of success occurs for $t_1 = \sqrt{2/3}$, $t_2 = \sqrt{3}/2$, and $t_3 = 1/\sqrt{2}$. This gives the maximum value of the coefficient $r_2 r_3$ that appears in Eq. (4), as can be seen in Fig. 6.

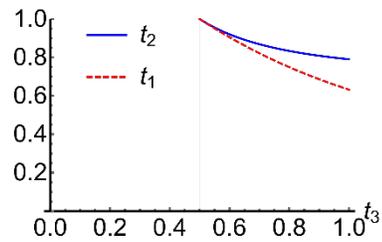

FIG 5. Transmission coefficients $t_1$ (dashed red line) and $t_2$ (solid blue line) that satisfy the necessary conditions $2t_1 t_2 t_3 = 1$ and $r_2 = r_1 t_2$, plotted as a function of $t_3$. These conditions are required for the successful operation of the destructive controlled-phase gate.

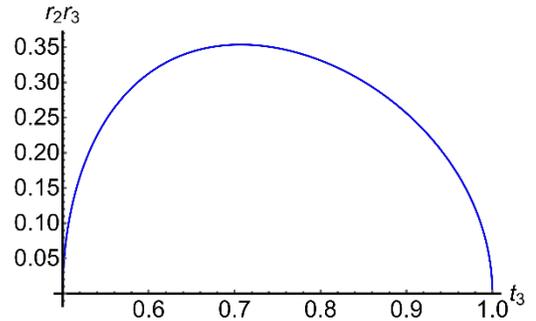

FIG 6. A plot of the maximum value of the coefficient $r_2 r_3$ as a function of the transmission coefficient $t_3$. Here $t_1$ and $t_2$ were chosen to satisfy the conditions the conditions $2t_1 t_2 t_3 = 1$ and $r_2 = r_1 t_2$ required for successful destructive controlled phase gate operation, as illustrated in Fig. 5. The maximum occurs at $t_3 = 1/\sqrt{2}$, which corresponds to using a 50-50 beam splitter in the photon subtraction.

From Eq. (4), the intrinsic probability $P_D$ of success of the destructive controlled-phase gate is given by

$$P_D = P_{NSG} \langle \psi' | \psi' \rangle$$
$$= P_{NSG} r_2^2 r_3^2 \left[ 1 + 2\left(|\alpha|^2 - |\beta|^2\right) \text{Re}\left(\gamma^* \delta\right) \right]. \quad (5)$$

Here $P_{NSG}$ is the probability of success for the nonlinear sign gate shown in Fig. 4. For the time being, we will assume that $P_{NSG}$ is calculated based on the assumption that the ancilla photons are produced with 100% efficiency.

$P_D$ depends on the values of the probability amplitudes $\alpha$, $\beta$, $\gamma$, and $\delta$ that describe the initial control and target qubits. This is illustrated in Fig. 7, which is a plot of the intrinsic probability of success as a function of $\alpha$ and $\gamma$, where all of the probability amplitudes were assumed to be real with $\beta = \sqrt{1-\alpha^2}$ and $\delta = \sqrt{1-\gamma^2}$, for example. In comparison, the probability of success is plotted in Fig. 8 for the case where the coefficients $\beta$ and $\delta$ were assumed to be imaginary instead. It can be seen that there is a significant variation in the probability of success depending on the form of the incident qubits.

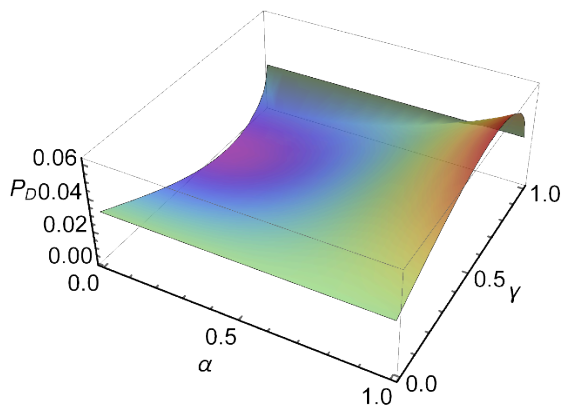

FIG 7. A plot of the intrinsic probability $P_D$ of success of the destructive controlled-phase gate as a function of the probability amplitudes $\alpha$ and $\gamma$ in the incident control and target qubits. All four probability amplitudes in Eq. (3) were assumed to be real in this example.

If the target qubit has a logical value of 1 ($\alpha = 0$) and $\gamma = \delta$, then it can be seen from Eq. (4) that the output state will have zero amplitude and $P_D = 0$, as can be seen in Fig. 7. This is an inherent feature of a destructive controlled-phase gate where the value of the control qubit is erased.

This does not occur for other values of the controlled phase shift, such as $\pi/2$, and it is not an issue in nonlocal interferometer applications, for example [3].

In order to simplify the comparison of the KLM controlled-phase gate and the gate proposed here, we averaged the intrinsic probability of success $P_D$ over all possible values of the coefficients $\alpha$, $\beta$, $\gamma$, and $\delta$. This result is compared with the corresponding result $P_{KLM}$ for the KLM controlled phase gate in Table 1. It can be seen that the intrinsic probability of success is comparable for the two gates for the case of $\phi = \pi$, which corresponds to a Controlled-Z operation.

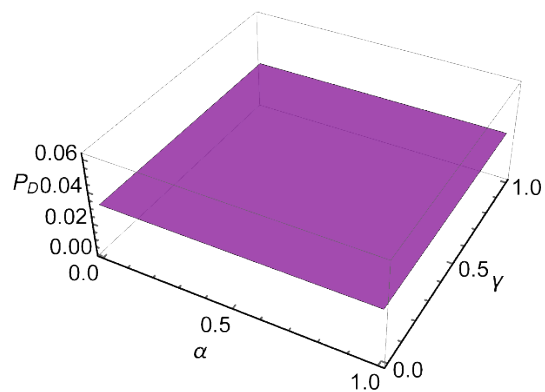

FIG 8. A plot of the intrinsic probability $P_D$ of success for the destructive controlled-phase gate as a function of $\alpha$ and $\gamma$, which were assumed to be real, while the remaining coefficients $\beta$ and $\delta$ were assumed to be imaginary.

Single photon ancilla can be generated using down-conversion and heralding on one of the pair of photons, which we will assume to succeed roughly 1% of the time [19]. Table 1 also includes the effective probabilities of success $P'_D$ and $P'_{KLM}$ for the two controlled phase gates if we include the probability of generating the required ancilla photons using down-conversion. It can be seen that $P'_D \gg P'_{KLM}$, since the KLM gate requires two ancilla photons while the destructive controlled-phase gate only requires a single longer ancilla photon.

| $P_D$ | 0.03125 |
|---|---|
| $P_{KLM}$ | 0.0625 |
| $P'_D$ | $3.125 \times 10^{-4}$ |
| $P'_{KLM}$ | $6.25 \times 10^{-6}$ |

**Table 1**. Comparison of the average probability of success of a destructive controlled-phase gate with that of a KLM gate, where $\phi = \pi$. Here $P_D$ and $P_{KLM}$ are the intrinsic success probabilities, while $P'_D$ and $P'_{KLM}$ include the probability of generating the required ancilla photons using heralded down-conversion.

As described in the previous section, a destructive controlled-phase shift of $\phi = \pi/2$ can also be produced using a different set of parameters. The KLM gate can be modified to produce a phase shift of $\phi = \pi/2$ as well [7]. The probability of success for these two gates was calculated in the same way as before and the results are compared in Table 2. It can be seen that the destructive controlled-phase gate has a much higher average probability of success in this case as well if we include the probability of generating the required ancilla photons using down-conversion and heralding.

| $P_D$ | 0.0226 |
|---|---|
| $P_{KLM}$ | 0.0327 |
| $P_D'$ | $2.26 \times 10^{-4}$ |
| $P_{KLM}'$ | $3.27 \times 10^{-6}$ |

**Table 2.** Comparison of the average probability of success of a destructive controlled-phase gate with that of a KLM gate for the case of a controlled phase shift of $\phi = \pi/2$.

In principle, both types of gates can be operated with 100% fidelity if the single-photon detectors are assumed to be perfect. The dark counts in an avalanche-diode single-photon detector are typically on the order of 100 counts/second or less. With a coincidence window of 1 ns, this corresponds to an erroneous output in approximately $10^{-7}$ of the events, which has a negligible effect on the fidelities.

In contrast, heralding on those cases where the output of a single-photon detector indicated that no photons were present can have a significant impact on the gate fidelity if the efficiency $\eta$ of the detectors is limited. Roughly speaking, this allows photons to escape unnoticed from the system, leaving an incorrect number of photons in the output state. The average fidelity $F_D$ of the destructive controlled-phase gate of Fig. 4 and the average fidelity $F_{KLM}$ for the KLM controlled-phase gate are plotted in Fig. 9 as a function of the detector efficiency $\eta$. Both of these results correspond to a controlled phase shift of $\phi = \pi$ and they assume that the ancilla photons have 100% fidelity.

It can be seen that the fidelity of the destructive controlled-phase gate is somewhat less than that of the KLM gate. This can be understood from the fact that the destructive controlled-phase gate of Fig. 4 relies upon 3 photon detectors indicating that no photons were detected, while the KLM gate of Fig. 3 only depends on 2 null detection events. This includes the fact that each of the nonlinear sign gates of Fig. 1 relies on a single null detection event.

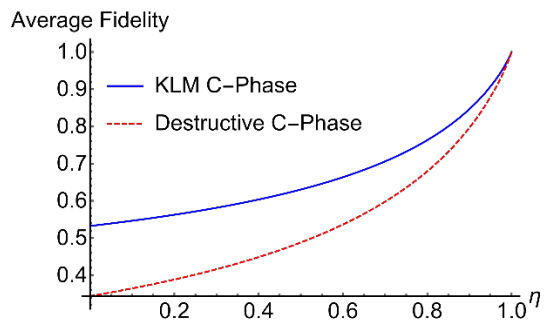

FIG. 9. Average fidelity $F_{KLM}$ of the KLM controlled-phase gate (solid blue line) compared with the average fidelity $F_D$ of a destructive controlled-phase gate (dashed red line). Both fidelities are plotted as a function of the single-photon detector efficiency $\eta$.

The KLM gate preserves the control qubit whereas it is destroyed in the controlled-phase gate of Fig. 4. As noted previously, a destructive controlled-phase gate can be used in a number of applications, such as nonlocal quantum interference experiments, the generation of entangled Schrodinger cat states [4], and in fusion operations for generating cluster states [11]. More generally, a quantum encoder [17] could be used in combination with a destructive controlled-phase gate to preserve the value of the control qubit, but that would require an additional ancilla photon. In that case, there would no longer be any advantage in the overall probability of success as compared to using the KLM gate.

## VI. Controlled phase shift for large photon numbers

Up to now, we have assumed that the target state that is input to the controlled-phase gate of Fig. 4 contains a maximum of one photon. There are potential applications where it would be desirable to produce a controlled phase shift on a state containing a larger number of photons, such as a coherent state. This can be useful in producing Schrodinger cat states [2] or in quantum interference experiments, for example [3,4].

The controlled-phase gate can be modified as shown in Fig. 10 to allow a larger number $n$ of photons in the input. Here a series of beam splitters is used to divide the incident field into $N$ different paths. For $N \gg n$, each of these paths will contain at most a single photon with high probability, which allows a destructive controlled-phase gate to be applied in each of the paths. The output of each of these controlled-phase gates can then be recombined using another series of beamsplitters. This approach is similar to the technique used for noiseless amplifiers when the input state has more than one photon [20].

The main limitation in this approach is that all of the controlled-phase gates have to succeed simultaneously, and the probability of that occurring decreases exponentially with the value of $N$. In addition, a single control qubit would need to control the phase shift in all $N$ paths. This can be accomplished by using a series of quantum encoders [10], which would further decrease the overall success rate. Nevertheless, an approach of this kind may be feasible for relatively weak coherent states.

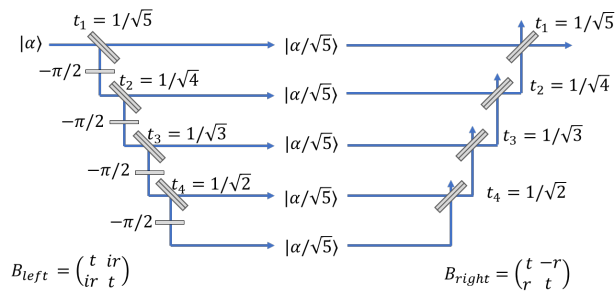

FIG. 10. A controlled-phase operation performed on an input state containing more than one photon, such as a coherent state. The incident field is divided into $N$ separate paths, each of which contains a destructive controlled-phase gate. The case of $N = 5$ is shown here. A set of beam splitters then recombines the individual beams to form a single output state.

## VII. Summary

We have proposed a destructive controlled-phase gate that produces a phase shift of $\phi$ when the control and target qubits both have a logical value of 1. The most commonly used values of $\phi$ are $\pi$ or $\pi/2$, but other phase shifts can be produced as well. The controlled-phase gate proposed here only requires a single nonlinear sign gate as a resource, whereas earlier implementations required two nonlinear sign gates [7]. As a result, the average probability of success for this controlled-sign gate is much larger than in earlier implementations if we include the need to generate ancilla photons using down-conversion and heralding. No such advantage would exist if the ancilla photons are produced on demand using quantum dots, but that typically does not give fidelities as high as can be achieved using down-conversion due to charge fluctuations [21]. Nevertheless, the use of quantum dots to produce single photons is an active area of research with continual improvements [22-24].

The basic idea behind the proposed controlled-phase gate is the use of a dual-rail control qubit to add a photon either before or after the nonlinear sign gate. If the photon is added before the nonlinear sign gate and the target qubit has a logical value of 1, then two photons will pass through the nonlinear sign gate and a phase shift of $\pi$ will be produced. No such phase shift will be produced if the photon addition is done after the nonlinear sign gate. A photon subtraction is performed at the output of the gate to restore the original number of photons in the target qubit.

The increased probability of success comes at the cost of destroying the control qubit. This is acceptable in a number of applications where the control qubit would have been destroyed in any event, such as in a postselection process. Potential applications of this kind include the generation of Schrodinger cat states [2], nonlocal interference experiments that violate Bell's inequality [4], and the construction of cluster states using fusion gates [11]. The control qubit can always be preserved if necessary by using a quantum encoder circuit [10] before the controlled-phase gate, but that would require two ancilla photons and there would be no benefit as compared to the original KLM controlled-phase gate. The probability of success vanishes for certain input states for a controlled phase of $\pi$, but that is not the case for other values of the controlled phase that are required in many applications.

In summary, the controlled-phase gate described here provides an interesting example of the use of photon addition and subtraction [12], and it may be of practical use in certain applications such as the generation of Schrodinger cat states and violations of Bell's inequality.

**Acknowledgements**

We would like to acknowledge valuable discussions with Cory Nunn and Todd Pittman. This work was supported in part by the National Science Foundation under grant number PHY-1802472.